\documentclass[aip,%
               apl,%
			   amsmath,%
			   amssymb,%
			   longbibliography]{revtex4-1}

\usepackage[separate-uncertainty,%
		    detect-all,%
            exponent-product=\cdot]{siunitx}
\usepackage{graphicx,nameref}
\usepackage{hyperref} 

\begin{document}

\author{Carlo Barth}
\author{Sebastian Roder}
\affiliation{Helmholtz-Zentrum Berlin f\"{u}r Materialien und Energie, %
	Kekul\'{e}str.~5, \mbox{12489 Berlin,~Germany}}
\author{Daniel Brodoceanu}
\author{Tobias Kraus}
\affiliation{Leibniz-Institut f\"{u}r Neue Materialien gGmbH, Campus D2 2, %
	\mbox{66123 Saarbr\"{u}cken,~Germany}}
\author{Martin Hammerschmidt}
\author{Sven Burger}
\affiliation{Zuse Institute Berlin, Takustr.~7, 14195 Berlin, Germany}
\author{Christiane Becker}
\email{christiane.becker@helmholtz-berlin.de}
\affiliation{Helmholtz-Zentrum Berlin f\"{u}r Materialien und Energie, %
	Kekul\'{e}str.~5, \mbox{12489 Berlin,~Germany}}

\noindent
{\small This manuscript was accepted by Appl.~Phys.~Lett.\newline
The citation for the published article is {\it C.~Barth, et al., Appl.~Phys.~Lett. {\bf 111}, 031111 (2017).}\\
Click here to see the published version: \url{http://dx.doi.org/10.1063/1.4995229}.}\\

\title{Increased Fluorescence of PbS Quantum Dots in
	Photonic Crystals by Excitation Enhancement}

\begin{abstract}
	We report on enhanced fluorescence of lead sulfide quantum dots
	interacting with leaky modes of slab-type silicon photonic crystals. The
	photonic crystal slabs were fabricated supporting leaky modes in the near
	infrared wavelength range. Lead sulfite quantum dots which are resonant the same
	spectral range were prepared in a thin layer above the slab. We selectively
	excited the leaky modes by tuning wavelength
	and angle of incidence of the laser source and measured distinct resonances of
	enhanced fluorescence. By an appropriate experiment design, we ruled out
	directional light extraction effects and  determined the
	impact of enhanced excitation. Three-dimensional numerical simulations
	consistently explain the experimental findings by strong near-field enhancements
	in the vicinity of the photonic crystal surface. Our study provides a basis for
	systematic tailoring of photonic crystals used in biological applications such
	as biosensing and single molecule detection, as well as quantum dot solar cells
	and spectral conversion applications.
\end{abstract}

\maketitle


\noindent Excitation and emission of fluorescent species can largely be affected
by leaky modes of photonic crystals (PhCs). PhCs can be designed with
specifically tailored optical properties\cite{Joannopoulos2008} and have been
intensely studied to improve systems that feature light emission, such as
light-emitting diodes (LEDs)\cite{Fan1997,Wiesmann2009} and
biosensors\cite{Cunningham2016}. Especially in the life-sciences, the
applications range from photonic crystal enhanced microscopy and single molecule
detection to enhanced live cell imaging, DNA sequencing and gene
expression\cite{Cunningham2016,Block2009,Ganesh2008a,Threm2012}. Two processes
can increase the light yield: (1) \emph{excitation enhancement} by generating
increased near-fields in the absorption wavelength range of the emitters, and
(2) \emph{extraction enhancement} by providing leaky-mode channels to improve
the out-coupling of the fluorescent light\cite{Boroditsky1999}.

Extraction enhancement has not only been demonstrated for LEDs, but for
nitrogen-vacancy centers in diamond\cite{Ondic2011}, DNA
microarrays\cite{Block2009}, molecules\cite{Ganesh2008,Mathias2008} and quantum
dots (QDs)\cite{Ryu2001,Ondic2012,Ondic2013,Ganesh2007}. Inducing extraction
enhancement with PhCs has been demonstrated several times, both for intrinsic
fluorescence of the PhC itself\cite{Ondic2011a,Ondic2011} and for QDs embedded
into the PhC\cite{Ganesh2008,Ondic2013} or on the PhC surface\cite{Ondic2012}.
Extraction enhancement effects cause strong directional features of the emitted
radiation which depend on the scattering angles of the involved leaky
modes\cite{Ganesh2008,Ondic2012}.

By excitation enhancement a significant increase in fluorescence of emitting
species is expected in the vicinity of the PhC surface since the near-field
energies of leaky modes can be orders of magnitude higher
compared to the incident field\cite{Becker2014a}. In this way, PhC structures
were applied to increase the efficiency of quantum dot solar
cells\cite{Adachi2013,Kim2013} and are shown to increase the quantum yield of
up-conversion devices\cite{Liu2013,Zhang2010a,Hofmann2016}. PhC-induced
excitation enhancement of fluorescent QDs was shown for individual
 modes or overlaid with directional extraction enhancement
effects\cite{Ganesh2007,Ganesh2008a}. However, an identification of
 modes exhibiting strong near-field energy
densities and the knowledge of the exact 3D spatial distribution in a defined
volume is of great importance for enhanced light excitation and emission
independent of directional characteristics, e.g. in biosensing, quantum dot
solar cells and up-conversion devices.

\begin{figure*}[t]
	\centering \includegraphics{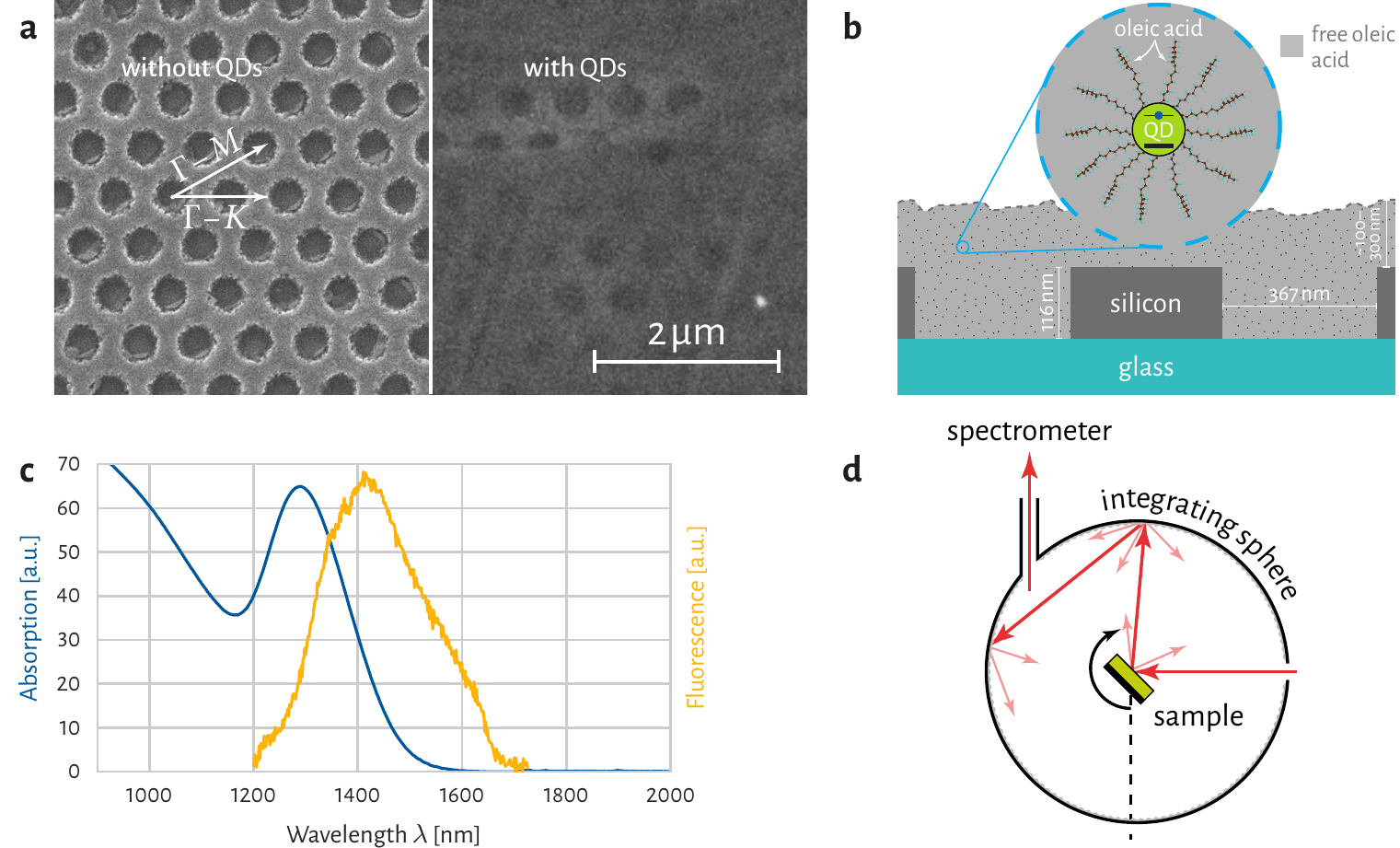} \caption{Sample
		details and experimental setup. (a) Scanning electron microscope image of the
		silicon photonic crystal without (left) and with (right) quantum dots. Arrows
		represent directions of the high-symmetry directions of the Brillouin zone. (b)
		Sketch of QDs in an oleic acid ligand on the surface of the photonic crystal
		(thickness of the QD containing layer not to scale). (c) Measured absorption of
		QDs  and a typical photoluminescence peak of QDs on a
		planar silicon sample.  (d) Sketch of the fluorescence measurement setup (see
		also Supporting Materials).} \label{fig:methods}
\end{figure*}

In this study, we experimentally map the integrated omnidirectional fluorescence
of lead sulfide (PbS) QDs interacting with leaky modes of a silicon photonic
crystal slab by using an angle- and excitation wavelength-resolved setup.
Directional characteristics of light extraction are eliminated by placement of
the sample inside an integrating sphere, hence allowing a pure determination of
excitation enhancement effects. By using finite element simulations, the
spectral position of the modes, as well as respective enhancement factors of the
electric field energy density are calculated. The comparison of experiment and
simulations enables us to explain the experimental fluorescence enhancement
features by leaky modes exhibiting a large local electric field energy density.

In order to realize a platform with increased electromagnetic near-fields, a
silicon PhC slab on glass with a hexagonal lattice of cylindrical air holes is
prepared (pitch: \SI{600}{\nm}, layer thickness: \SI{116}{\nm}, hole radius:
\SI{367}{\nm}) using a process based on nanoimprint lithography, physical vapor
deposition of silicon and solid phase crystallization\cite{Becker2014a,
Barth2016}. Figure~\ref{fig:methods}(a, left) shows a scanning electron
microscope image of the PhC with the two polar angle directions corresponding to
the $\Gamma-M$ and $\Gamma-K$ paths between the high-symmetry points of the
irreducible Brillouin zone. Subsequently, the PhCs are covered with spherical
PbS QDs with a diameter of \SI{5.6(8)}{\nm} (manufacturer: CAN GmbH) using the
convective assembly technique\cite{Kraus2007} -- a high accuracy process that
distributes the QDs evenly across the sample (Fig.~\ref{fig:methods}(a, right)).
They are originally solved in toluene and their surface is shielded by an oleic
acid ligand (the toluene is removed in the covering process).
We note that we value the thickness of the QD-oleic acid layer based
on SEM images to range from about $\SI{100}{\nm}$ to about $\SI{300}{\nm}$.
Figure~\ref{fig:methods}(b) shows a sketch of the complete system in cross
section after the particle assembly process. The absorption properties (blue) of
the QDs as well as the fluorescence peak (orange) are shown in
Fig.~\ref{fig:methods}(c). These QDs can be pumped efficiently in a large
spectral range from the visible up to about \SI{1500}{\nm}, while the
fluorescence peak is centered at about \SI{1400}{\nm}. The fluorescence is
measured by placing the samples inside an integrating sphere with diameter
\SI{5}{\cm} (Fig.~\ref{fig:methods}(d)). For excitation a tunable laser source
with a wavelength range of \SI{1080}{\nm} to \SI{1140}{\nm} is used (SACHER Lion
Series) with linear polarization and a line width of $\sim\SI{3}{\nm}$. A
fiber-coupled output port guides the light to the spectrometer (Ocean Optics)
after passing an edge filter to suppress the excitation laser radiation. In
order to vary the incident angle the sample is rotated inside the sphere. For
further details on the experimental setup please see Fig.~S1 in the
Supplementary Material. Due to the usage of an integrated sphere the
fluorescence measurement is independent of directional effects of the light
emitted by the quantum dots. We have further checked that resonantly enhanced
emission does not play a quantitative role in our experimental setup by
measuring the spectral distribution of emitted light: this spectrum does not
depend on the presence of the PhC.

\begin{figure*}[t]
	\centering \includegraphics[scale=1.11]{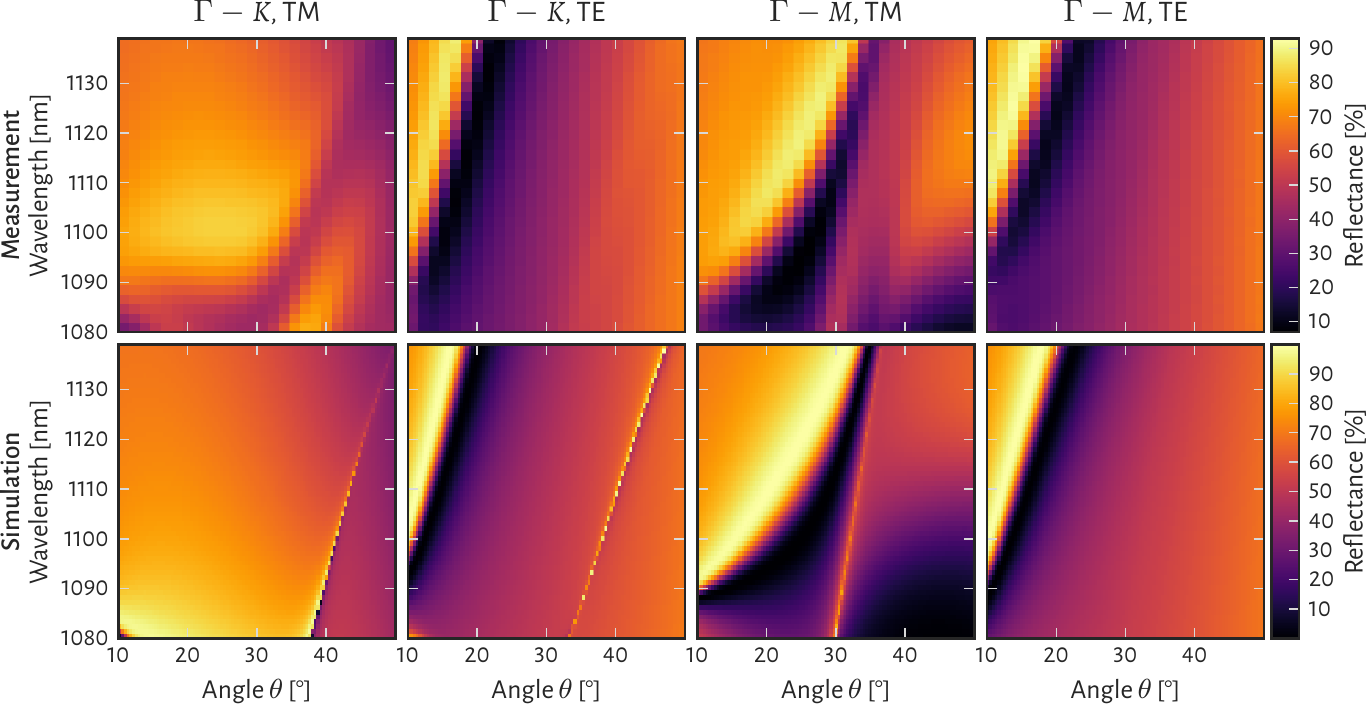}
	\caption{Comparison of experimental and numerical reflectance of the photonic
		crystal without quantum dot coating. The experimental reflectance properties
		obtained from angular resolved reflectance measurements (upper row) are
		compared to simulation results in the same wavelength- and angle-range after
		geometry reconstruction (lower row).} \label{fig:ARR_vs_RB_FEM}
\end{figure*}

The fluorescence of the PbS QDs can be enhanced by an interaction with the leaky
modes of the PhC. Therefore, leaky modes of the PhC were identified by angular-
and wavelength-resolved reflectance measurements using a PerkinElmer Lambda 950
spectrometer with automated reflectance/transmittance analyzer supplement.
Resonant features in the reflectance spectra can be attributed to the coupling
of incident light to the photonic bands of the PhC\cite{Astratov1999}.
Figure~\ref{fig:ARR_vs_RB_FEM} (upper row) shows the measured reflectance
properties of the silicon PhC slabs for the four combinations of sample
orientation (corresponding to $\Gamma-M$ and $\Gamma-K$) and source polarization
(transversal electric, TE, and transversal magnetic, TM).  We observe
reflectance values of almost \SI{0}{\percent} up to more than \SI{90}{\percent},
indicating the presence of photonic bands. We decided to use a wavelength range
of \SIrange{1080}{1140}{\nm} for the excitation (marked in
Fig.~\ref{fig:ARR_vs_RB_FEM}). In this range, (1) the QDs show a considerable
absorption, (2) do not show a fluorescence themselves and (3) the absorption of
the silicon is low -- which would deteriorate near-field enhancement
effects\cite{Becker2014a}. Notably, there are only a few well-isolated bands in
the excitation wavelength regime which facilitates the interpretation of
subsequent fluorescence measurements. For simulations we use a time-harmonic
finite-element Maxwell solver (JCMsuite\cite{Pomplun2007}) on a 2D-periodic
unstructured, prismatoidal mesh of the unit cell of the hexagonal sample
structure consisting of a glass subspace and a silicon-hole-layer on top. Each
simulation uses a distinct plane wave excitation corresponding to the direction
of incidence, wavelength and polarization. We assure numerical convergence by
comparing the derived quantities to those calculated in a highly accurate
reference solution, guaranteeing an accuracy of \SI{1}{\percent}. Optical
constants are taken from tabulated values\cite{Palik1985} for glass and silicon,
considering all the significant physical effects including dispersion and
absorption. The numerically obtained reflectance maps calculated using Fourier
transforms on the required domain boundaries are shown in the lower row of
Fig.~\ref{fig:ARR_vs_RB_FEM}. A comparison with the measured reflectance maps
allowed us to reconstruct the exact geometrical parameters of the produced
sample using the reduced basis method\cite{Hammerschmidt2016}. While a first
guess from experimental scanning electron microscopic images yielded a silicon
thickness of $\sim\SI{130}{\nm}$ and a hole diameter of $\sim\SI{360}{\nm}$, the
numerical reconstruction generated more exact values with a silicon thickness of
\SI{116}{\nm}, a hole diameter of \SI{367}{\nm} and a lattice pitch of
\SI{600}{\nm}. The maps agree excellently with the measured results in the whole
spectral range. We reason from this comparison that the numerical model gives
accurate predictions for the fabricated sample.

\begin{figure*}[t]
	\centering \includegraphics{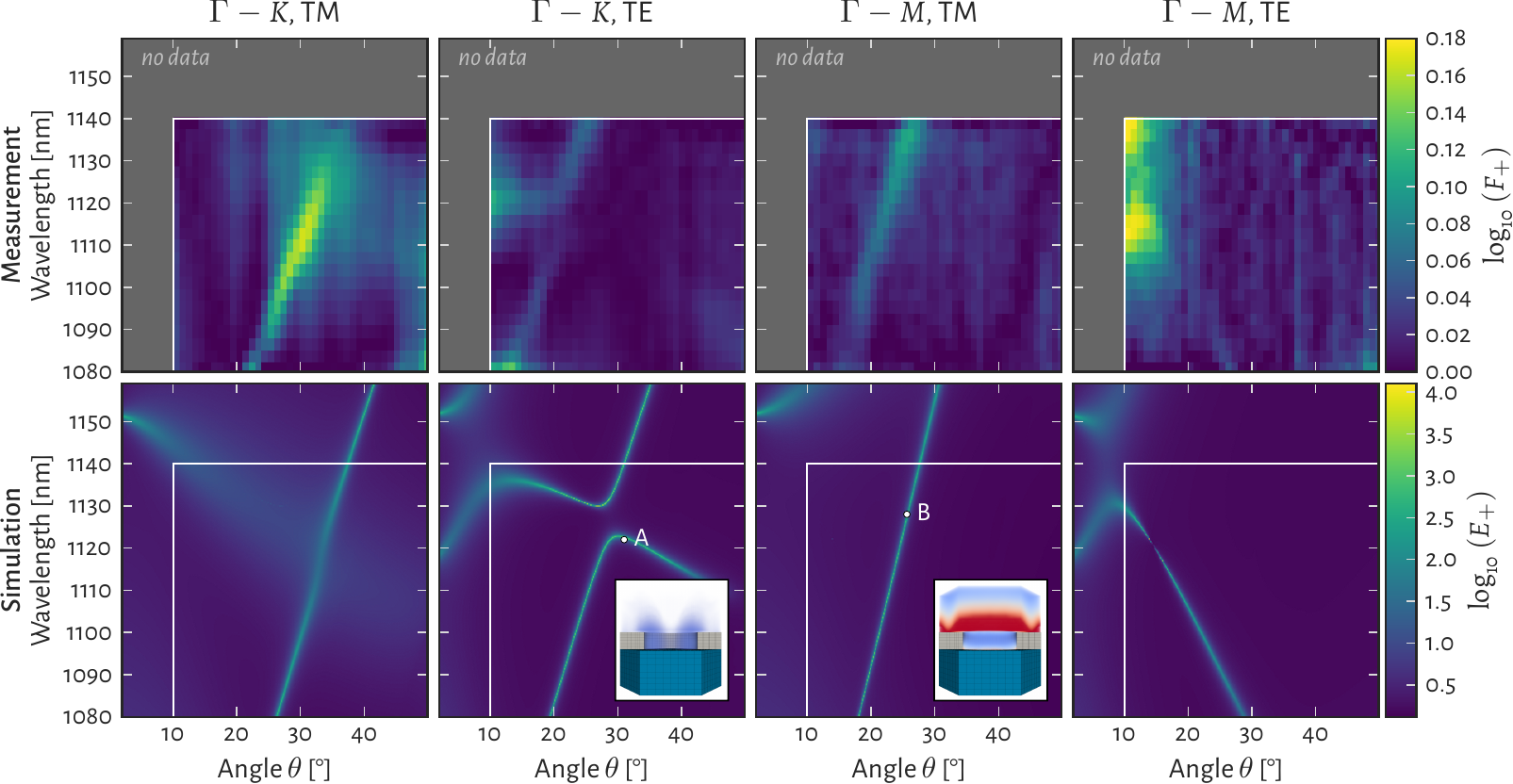} \caption{Measured
		fluorescence enhancement maps (upper row) and simulated electric field energy
		enhancement maps (lower row). The insets show 3-dimensional field energy
		distributions for the two selected points A and B. The white lines in the
		simulation data mark the experimental data limits.}
	\label{fig:fluorescence_vs_FEM}
\end{figure*}

We then measured the fluorescence of the PbS-QDs on the PhC surface
experimentally in an angle- and wavelength-resolved setup
(Fig.~\ref{fig:methods}(d)) in a region of high QD-layer thickness
in order to reach a sufficient signal (i.e. $\sim\SI{300}{\nm}$). 
To achieve the fluorescence enhancement maps
depicted in Fig.~\ref{fig:fluorescence_vs_FEM} (top row), each measured spectrum
was first integrated over the fluorescence peak from $\lambda=\SI{1200}{\nm}$ to
$\lambda=\SI{1700}{\nm}$ and normalized to the measured incident laser power and
the absorption profile of the quantum dots. We further characterized the
dependence on the incident angle of the integrating sphere using a planar
reference (i.e. the PbS/oleic acid system on an unpatterned silicon film of the
same thickness as the PhC slab) and corrected the results correspondingly. With
these corrections and subsequent background filtering, we find maps for the
actual fluorescence $F$ of the QDs. We find a minimal estimate for the
fluorescence enhancement $F_+$ shown in Fig.~\ref{fig:fluorescence_vs_FEM} (top
row) by dividing by the minimal value in each of the measurement windows. We
observe clear and sharp features of enhanced fluorescence with a maximum
enhancement factor of $F_+=\num{e0.18}\approx \num{1.5}$, i.e. an increase of
roughly \SI{50}{\percent}. An isolated steep band of enhanced fluorescence is
observed for TM polarized light. For TE polarized light also distinct features
appear in the fluorescence enhancement maps.

For a better understanding, we compare these experimentally determined
fluorescence results to numerically obtained field energy enhancement maps in
the bottom row of Fig.~\ref{fig:fluorescence_vs_FEM} (using the same logarithmic
color scale for all maps). Please note that upper and lower row do not show the
same physical quantity. For the determination of the electric field energy
enhancement $E_+$, we integrate the electric field energy density
distribution $u(\mathbf{r})=\tfrac{1}{4}\mathbf{D}\cdot\mathbf{E}$
over the superspace volume $V$ -– defined by the PhC hole and a \SI{250}{\nm}
layer above the silicon -- and normalize it to the energy of the incident plane
wave source in the same volume $U_\mathrm{pw}=\epsilon_0 V|\mathbf{E}|^2/4$. We
numerically treat the PbS/oleic acid system in this superspace volume as an
infinite space of an average refractive index material with
$n_\mathrm{superspace}=1.65$, estimated by thermo gravimetric mass fraction
analysis and the respective densities and refractive indices of both materials
in the wavelength range of interest ($n_\mathrm{PbS}\approx 4.3$,
$n_\mathrm{oleic acid}\approx 1.5$). In the simulations, we observe field energy
enhancement values of up to \num{10000} correlated with peak width (FWHM) of
\SI{0.3}{\nm}. Such a resolution is not reached in the experiment due to the
laser bandwidth of \SI{3}{\nm}. Geometrical imperfections of the PhC further
broaden the leaky mode resonances. This partly explains why large simulated
electric field enhancement factors of up to \num{10000} only result in
fluorescence enhancements of \SI{50}{\percent}. A further reason might be the
fact, that the thickness of the QD containing layer is much larger than the
spatial extend of the  modes. Exemplary, two numerical field
energy distributions are shown as insets in Fig.~\ref{fig:fluorescence_vs_FEM}.
The leaky mode (A) is located very close to the PhC surface such that only a
small fraction of the quantum dots located closely on the photonic crystal
surface is exposed to the enhanced fields. In addition, the
corresponding peak has a high Q-factor of $\sim\num{997}$ (i.e. a FWHM of
$\approx\SI{1.12}{\nm}$) which causes an ineffective excitation with a laser
bandwidth of $\SI{3}{\nm}$ (see the supplementary material, Tab.~S1, for
additional Q-factors and discussion). Hence, no fluorescence enhancement is
visible in the experiment. However, resonance (B) exhibits a much larger extent
and overlap with the QDs which can be clearly correlated with a measurable
fluorescence enhancement. This is also supported by the Q-factor of
$\sim\num{396}$ (FWHM$\approx\SI{2.85}{\nm}$), which fits the laser bandwidth
much more accurately.  A larger amount of simulated field energy distributions
is shown in Fig.~S2 of the supplementary material. Nevertheless, the qualitative agreement of
the position of the resonances regarding wavelength and angle of incidence is
excellent: note the almost isolated steep band of high field
energy enhancement occurring in the two maps with TM polarized light, which is
also clearly visible in the measurement with a very good accordance in their
gradient angles and spectral positions. The TE-cases feature bands that show
anti-crossing behavior in the simulations\cite{Barth2016} which can also partly
be observed in the experiment. We explain the distinct differences between
experimental and numerical results in the TE-cases by Q-factor
mismatches and the above mentioned insufficient overlap of leaky mode and QDs,
which is in fact obvious from a 3D field energy distribution analysis.

In conclusion, we measured increased fluorescence by excitation enhancement of
PbS QDs on the surface of a hexagonal-lattice silicon PhC slab with a maximum
enhancement of \SI{50}{\percent}. Directional extraction characteristics were eliminated by an appropriate
experimental setup allowing for an undisguised determination of excitation
enhancement effects. We explained our findings by the interaction of the QDs
with leaky modes exhibiting large field energy enhancement values close to the
PhC surface. A numerical analysis yielded the 3-dimensional distribution of the
involved leaky modes and their spatial overlap with the QDs well explaining the
presence or the absence of enhanced fluorescence. We conclude that a systematic
placement of emitting species directly on the photonic crystal surface
enables to increase the fluorescence
enhancement factors. The knowledge of strength and spatial distribution of light
interacting with PhC leaky modes will be of importance for excitation enhanced
future devices in the fields of biosensors, photochemistry, up-conversion and QD
solar cells.

\section*{Supplementary material} 

See supplementary material for details on the fluorescence enhancement
measurement setup, additional simulated 3-dimensional
electric field energy distributions and a Q-factor analysis.

\begin{acknowledgments}

The authors thank David Eisenhauer, Grit K{\"{o}}ppel, Bernd Rech and Klaus
J{\"{a}}ger from Helmholtz-Zentrum Berlin for support at sample fabrication and
useful discussions. The German Federal Ministry of Education and Research is
acknowledged for funding the research activities of the Nano-SIPPE group within
the program NanoMatFutur (No. 03X5520). Further we acknowledge support by the
Einstein Foundation Berlin through ECMath within subproject OT9. Parts of the
results were obtained at the Berlin Joint Lab for Optical Simulations for Energy
Research (BerOSE) of Helmholtz-Zentrum Berlin f\"{u}r Materialien und Energie,
Zuse Institute Berlin and Freie Universit\"{a}t Berlin.

\end{acknowledgments}

%

\end{document}